\documentclass[prb,twocolumn,showpacs,floatfix]{revtex4-1}
\usepackage{graphicx}
\usepackage{color}
\usepackage{amssymb}
\usepackage{amsmath}
\usepackage{xspace}
\usepackage{url}
\usepackage{subfigure}
\usepackage{float}

\begin{document} 
\title{Dot-bound and dispersive states in graphene quantum dot superlattices}
%
\author{A. Pieper,$^1$ R. L. Heinisch,$^1$ G. Wellein$^2$ and H. Fehske$^{1}$}
\affiliation{$^1$Institut f\"ur Physik, Ernst-Moritz-Arndt-Universit\"at
  Greifswald, 17487 Greifswald, Germany}
\affiliation{$^2$Regionales Rechenzentrum Erlangen, 
Universit\"at Erlangen-N\"urnberg, 91058 Erlangen, Germany}
\date{\today}
\begin{abstract}
We consider a square lattice configuration of circular gate-defined quantum dots in an unbiased graphene sheet 
and calculate the electronic, particularly spectral properties of finite albeit actual sample sized systems
by means of a numerically exact kernel polynomial expansion technique. Analyzing the local density of states and the 
momentum resolved photoemission spectrum we find clear evidence for a series of quasi-bound states at 
the  dots, which can be probed by optical measurements. We further analyze the interplay of the superlattice structure with dot localized modes on the electron energy dispersion. Effects of disordered dot lattices are discussed too.
\end{abstract}
\maketitle

\section{Introduction}
Graphene, in which the carbon atoms are condensed in a strictly two-dimensional honeycomb lattice due to their $sp^2$ hybridization, 
constitutes a unique form of quantum matter,  interesting for both fundamental science and applications.\cite{NGMJZDGF04,CGPNG09,AABZC10,KUPGC12,BLHSCF12} 
Specifically the electronic properties of graphene are extraordinary. Graphene is a topological material where the  quasiparticles 
(low-energy excitations) near the so-called Dirac nodal points behave as massless relativistic Dirac fermions possessing a linear energy dispersion. 
In neutral graphene the Fermi energy crosses exactly the Dirac points. Hence, having a vanishing density of states at the Fermi energy but no gap in 
the excitation spectrum, the system is a hybrid between a metal and a semiconductor.  These features lead to many unusual and sometimes
counterintuitive transport phenomena such as a finite  universal dc conductivity at the neutrality point, Klein tunneling,  
or an anomalous quantum Hall effect.\cite{CGPNG09}

From an application-technological point of view, the tunability  of graphene's electronic and optical properties by external electric and magnetic fields is of particular importance.\cite{Go11} Amongst others, this provides unique possibilities to modify---for instance, by gating---the properties of finite areas in graphene-based structures.     
Exploiting the parallels between optics and (Dirac) electronics, this has led to the proposal of potential steps such as Veselago lenses for propagating electron beams\cite{CFA07} 
or the experimental implementation of the counterpart of optical fiber cables.\cite{WLLM11}
For a circular gated region refraction at the boundary  leads to two coalescing caustics that focus the electron density in the dot.\cite{CPP07} Similarly, circular gated regions have also been studied in bilayer graphene\cite{PPC09} and in monolayer graphene with spin-orbit coupling which induces birefringence.\cite{AU13} 
Interestingly, circular dots in unbiased graphene allow electrostatic electron confinement in spite of Klein tunneling.\cite{BTB09} For biased graphene long-living temporary bound states appear.\cite{HA08} These modes entail a peculiar angular scattering characteristic: In particular forward scattering and Klein tunneling can be almost switched off by a Fano resonance phenomenon.\cite{HBF13a,PHF13} While quantum dots on etched graphene have been studied as potential hosts for spin qubits,\cite{TBLB07,WRAWRB09,PSKYHNG08,RT10} single gate-defined dots\cite{HG07} and multiple dots arranged in corrals\cite{VAW11} have been used to model the scattering of Dirac electron  waves by impurities or metallic islands placed on a graphene sheet.
If there is more than a single graphene quantum dot, interdot coupling---realized, e.g., via direct tunneling between the dots or through the continuum states of graphene---gains in importance.\cite{HA09} 

In this contribution, we investigate the electronic properties of 
graphene gate-controlled quantum dots arranged in a regular square lattice configuration. Graphene superlattices  offer an exciting prospect to tailor the charge carrier behavior through a renormalization of the group velocity at the Dirac point\cite{} or the emergence of higher order Dirac points.\cite{PYS08,PYSCL08,PSYCL09,BVP10a,Poetal13}
Using the kernel polynomial method (KPM)\cite{WWAF06,WF08} in order to obtain unbiased numerical results, we monitor the (local) density of states (LDOS/DOS), the optical conductivity of the sample and the single-particle excitation spectrum.
Thereby, we first  discuss the existence of quasi-localized states at the quantum dot regions with energies near the normal-modes of an isolated free-standing graphene dot. Then we consider the interplay of the additional flat bands stemming from the normal modes of the dots with sublattice effects such as higher order Dirac cones and group velocity renormalization. Finally, we demonstrate how different types of disorder destroy or preserve superlattice and normal mode induced spectral signatures.

\section{Model and method}
\begin{figure}[t]
\centering
\includegraphics[width=0.55\linewidth]{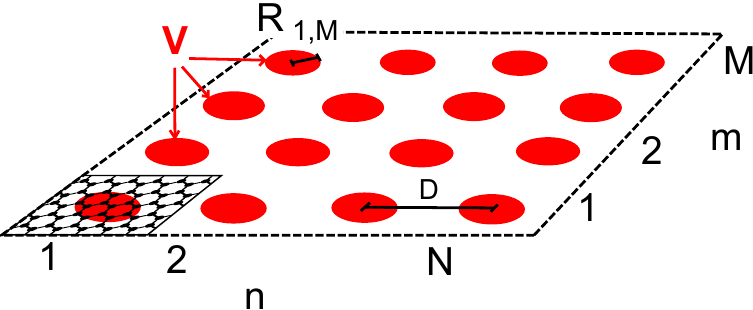}\hspace*{.5cm}\includegraphics[width=0.4\linewidth]{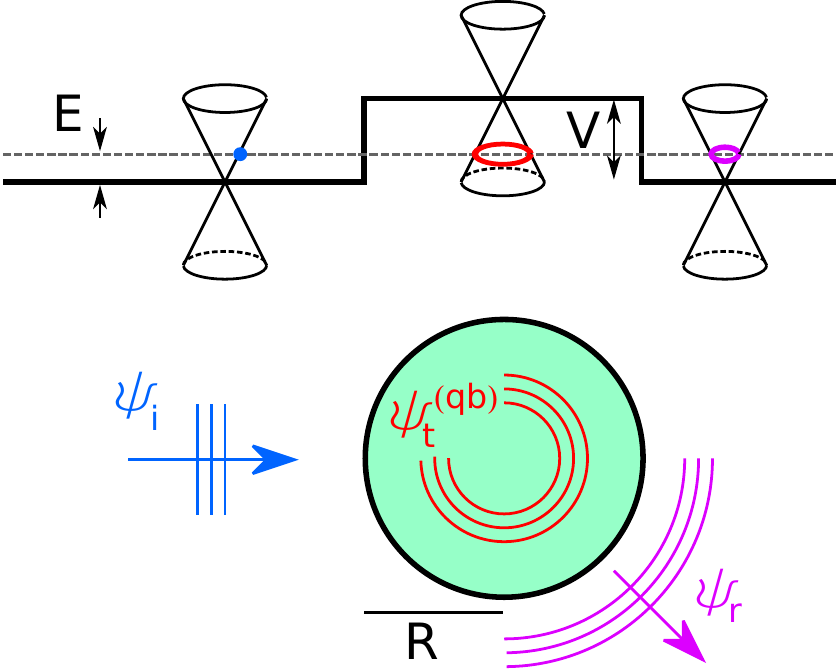}
\caption{(Color online) Left: Graphene quantum dot array used in this work. The dots are defined electrostatically,  by applying a constant bias $V$. The dot radius is $R$, 
the square dot superlattice constant is $D$.  To ensure the gate potential to be smooth on the scale of the lattice spacing $a$, we adopt a linear interpolation of $V_i$ within a small range 
$R \pm 0.01 R$. The underlying graphene honeycomb lattice structure is shown in the lower left corner;
we have zigzag (armchair) edges in $x$ $(y)$ direction with $N$ $(M)$ dots.  Periodic boundary conditions (PBCs) were used at the  edges of the sample. 
Right: Sketch of Dirac electron scattering  at a single quantum dot.  For $E<V$, where the dot embodies an n-p junction, the incident ($\psi_i$) and reflected  ($\psi_r$) electron waves reside in the conduction band, while the transmitted ($\psi_t^{(qb)}$) wave inside the dot corresponds to a state in the valence band. 
Owing to the  double-cone dispersion (near $K$ and $K'$) non-evanescent waves can exist in the dot, i.e., $\psi_t^{(qb)}$ might give rise to a quasi-bound state.}
\label{fig1}
\end{figure}

We model the electronic structure by a tight-binding Hamiltonian,
  \begin{equation}\label{H_bm}
     {H} =  \sum_{i}V_i^{} {c}_i^{\dag} {c}_i^{} 
           -t \sum_{\langle ij \rangle}({c}_i^{\dag} {c}_j^{} + \text{H.c.})\,,
  \end{equation}
where $c_i^{(\dag)}$ is a fermionic annihilation (creation) operator acting on lattice site $i$ of the honeycomb lattice with $L$ sites. The nearest neighbor hopping amplitude is $t\simeq$ 3~eV. The setup we consider (see Fig.~\ref{fig1}) is a graphene sheet on a gated substrate with circular regions where an additional external potential 
\begin{equation}
V_i=V \sum_{(n,m)} \Theta (R-|\vec{r}_i-\vec{r}_{(n,m)}|)
\end{equation}
is applied.

The local electronic properties of this graphene quantum dot superlattice are reflected in the LDOS, 
\begin{equation}
\rho_i(E)=\sum_{l=1}^L |\langle i | l\rangle |^2 \delta (E-E_l)\,,
\label{ldos}
\end{equation}
where $|i\rangle= c_i^\dagger |0\rangle$, and $|l\rangle$ is a single-electron eigenstate of $H$ with energy $E_l$. 
The LDOS can be directly probed by scanning tunneling microscopy.\cite{NKF09} For the noninteracting system~(\ref{H_bm}), 
$\rho_i(E)$ can be determined  to, {\it de facto}, arbitrary precision by the KPM, which is based on an expansion of the (rescaled) Hamiltonian 
into a finite series of Chebyshev polynomials.\cite{WWAF06,WF08} The mean DOS follows as $\rho(E)=\sum_{i=1}^L \rho_i(E)$. 

The momentum-resolved single-particle spectral function at zero temperature, 
  \begin{equation}\label{eq:A_k}
    {A(\vec k, E)} =  \sum_{n=1}^{L} | \langle l | \psi (\vec k) \rangle |^2 \delta (E - E_l)\,,
  \end{equation}
is easily accessible by the KPM as well.\cite{WWAF06,WF08}   Here $| \psi (\vec k) \rangle = L^{-1/2} \sum_i \exp(i\vec k \vec r_i) c_i^{\dag} | 0 \rangle$. Note that  $| \psi (\vec k) \rangle$ is not  a Bloch eigenstate of infinite graphene due to its sublattice structure.

Within our KPM scheme, we also have access to the real part of the optical conductivity:\cite{WWAF06,WF08} 
  \begin{align} \label{eq_reg_opt_Leitfahigkeit}
  \sigma(\omega) =& 
          \frac{\pi \hbar}{\omega\Omega}\sum_{l,l'} |\langle l |  J_x | l' \rangle |^2\,[f(E_l) - f(E_{l'})]\delta \left( \omega + E_l - E_{l'} \right)
  \end{align}
  with $ J_x=-(\text{ie}t/\hbar)\sum_{\langle i,j\rangle} (r_{j,x}-r_{i,x}) c_i^\dagger c_j^{} $
 the $x$-component of the current operator. In (\ref{eq_reg_opt_Leitfahigkeit}),  $f(E)= [e^{ (E - \mu)/T}+1]^{-1}$ denotes 
the Fermi function containing the temperature $T$ and the chemical potential $\mu$. Moreover, $\Omega=3^{3/2}La^2/4$, where $a\simeq 1.42$~\r{A} is the carbon-carbon distance. 

\section{Numerical results and discussion}
We begin our discussion with the signatures of localized modes for a graphene sample with a regular array of quantum dots. To identify dot-induced features, we compare with results for a single circular quantum dot with sharp boundary in an infinite graphene sheet treated within the continuum Dirac-equation 
approximation.\cite{MP08,HA08,BTB09,PAS11,HBF13a} In this case the electronic states in the dot are resonances with finite trapping time. 
Due to interference effects the trapping time may even become infinite in unbiased graphene for particular parameters of the sharp circular confinement potential.\cite{HA08,BTB09,HBF13a} 
The quasi-bound states for an isolated dot, $a_m$,  can be classified according to their angular momentum. The mode $a_m$ is made up of states with total angular momentum $j=\pm(m+1/2)$ (composed of orbital momentum $m$ and pseudo spin $\pm 1/2$). The modes $a_m$ are fourfold degenerate: twice with respect to $\pm j$ and twice with respect to valley degrees of freedom $K$ and $K'$.  For small energies the mode $a_0$ is relatively broad. While contributing significantly to electron scattering it does not evolve into a true bound state for $E\rightarrow 0$. For higher modes, however, the electron is strongly localized at the dot, overcoming Klein tunneling. For unbiased graphene dot-localized modes appear for the 'dot parameter' $\eta=VR/v_F=j_{m,s}$, where  $v_F=3at/2 \hbar$ is the Fermi velocity in pristine graphene and $j_{m,s}$ is the $s$th zero of the Bessel function $J_m$.\cite{BTB09} We note that electron confinement of such kind can persist for relatively small dots,  even taking the lattice discreteness into account.\cite{PAS11,PHF13}
If we consider an array of gate-defined quantum dots, for very large interdot  distances, $D\gg R$, all dots have the same energy spectrum. When $D$ comes up to $R$'s order of 
magnitude, the interdot coupling results in a splitting of the degenerate energy levels. This has been demonstrated for a periodic chain of quantum dots.\cite{HA09} 

\begin{figure}[t]
\centering
\includegraphics[width=0.98\linewidth]{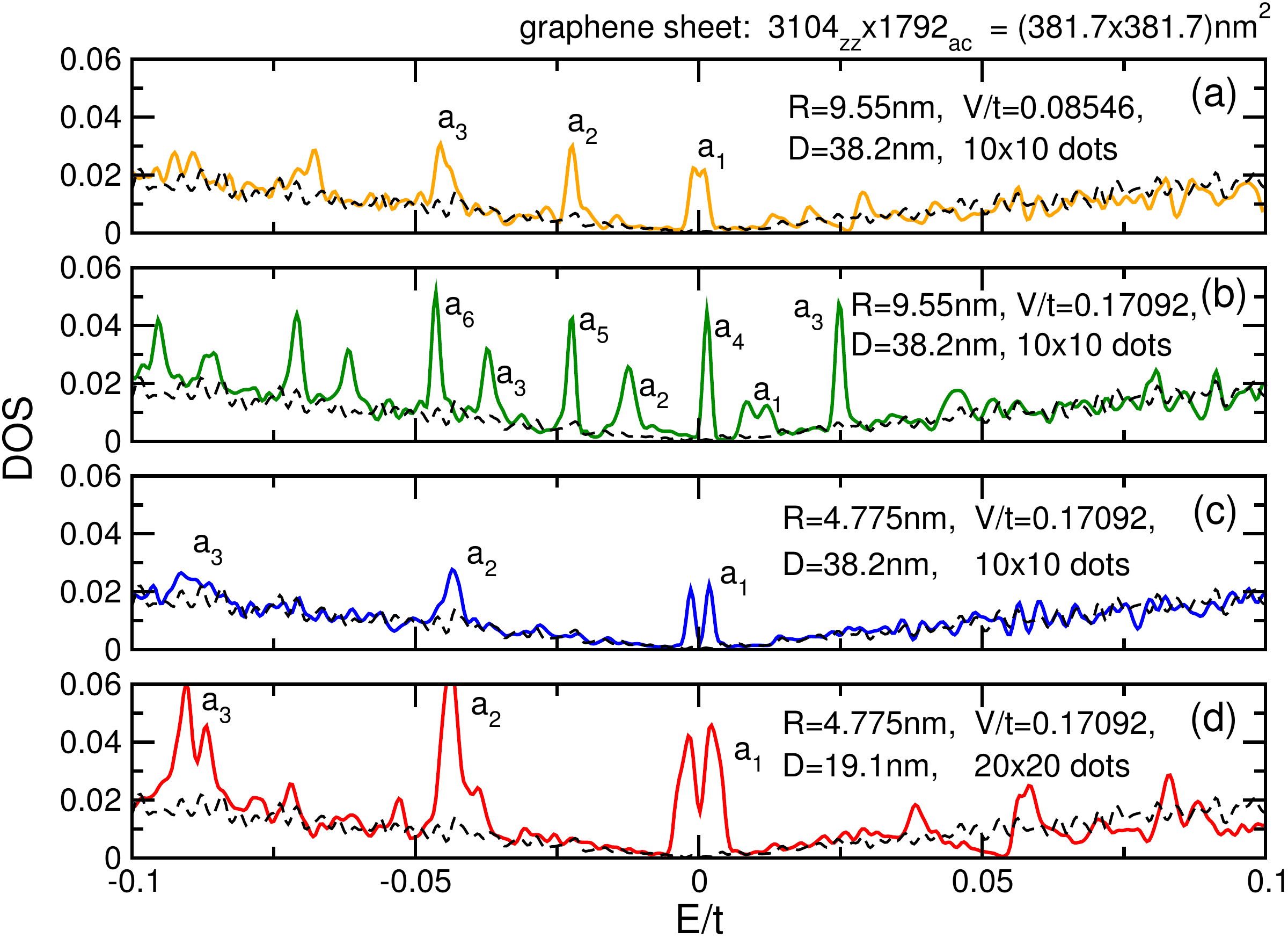} 
\caption{(Color online) DOS of the graphene quantum dot superlattice in dependence on $R$, $D$, and $V/t$.  Peaks related to quasi-bound dot modes are designated by $a_m$. Note that the dot parameter $\eta$ is the same in panels (a), (c) and (d). The DOS is calculated by the KPM on a lattice with 3104 (1792) sites in zigzag (armchair) edge direction, using 16384 Chebyshev moments. To identify effects due
to the finiteness of the sample, the PBCs and the resolution of the KPM---leading to small deviations from the strictly linear increase of the DOS 
near $E=0$---we included the DOS obtained numerically for the case $V_i=0$ $\forall i$ (dashed lines). The total DOS, of course, fulfills the sum rule $\int_{-\infty}^\infty \rho(E) dE=1$. }
\label{fig2}
\end{figure}

We now compare the DOS of samples with and without quantum dot superlattice (see Fig. \ref{fig2}). If all $V_i=0$, the DOS of the nearest-neighbor $\pi$ electron tight-binding model~(\ref{H_bm}), describing pure graphene in that case, can be calculated
analytically in terms of a complete elliptic integral of the first kind with energy-dependent prefactors.\cite{HN53} Most notably,
close to the Dirac points $K$ and $K'$, $\rho(E)$ is proportional to $|E|/v_F^2$. Figure~\ref{fig2} gives the DOS near the Dirac point of a finite graphene system with PBCs, where 
10$\times$10 [panels (a)-(c)] respectively 20$\times$20 [panel (d)]  quantum dots were 
arranged periodically in a square (super-) lattice configuration. Contributions emanating from quasi-bound modes $a_m$ are superimposed on the DOS of pristine graphene. They form narrow energy bands, except for the broad $a_0$ mode.  
 In panel (a) the voltage $V/t=0.08546$ was chosen to fix the lowest $a_1$ mode---originally located in the lower Dirac cone---at zero energy.  
 If we further increase the gate potential the energy-ladder of dot bound-states is shifted upwards. Accordingly, in panel (b), where $V/t=0.17092$, the first $a_4$ related band 
 has reached the Fermi energy, whereas  states assigned to the first and second $a_1$, as well as to the first $a_2$ and $a_3$ resonances  passed the Dirac point already.   
 We note that bands belonging to $a_m$ states with larger $m$ are less spread in energy. Panel (c) shows that reducing the size of the quantum dots the different quasi-bound states 
 become more separated energetically.  Finite dot-size effects provoke the splitting of some $a_m$ bands.  A larger number of 
 dots will of course enhance the weight of the quasi-bound states [compare panels (c) and (d)].  

\begin{figure}[t!]
\centering
\includegraphics[width=0.98\linewidth]{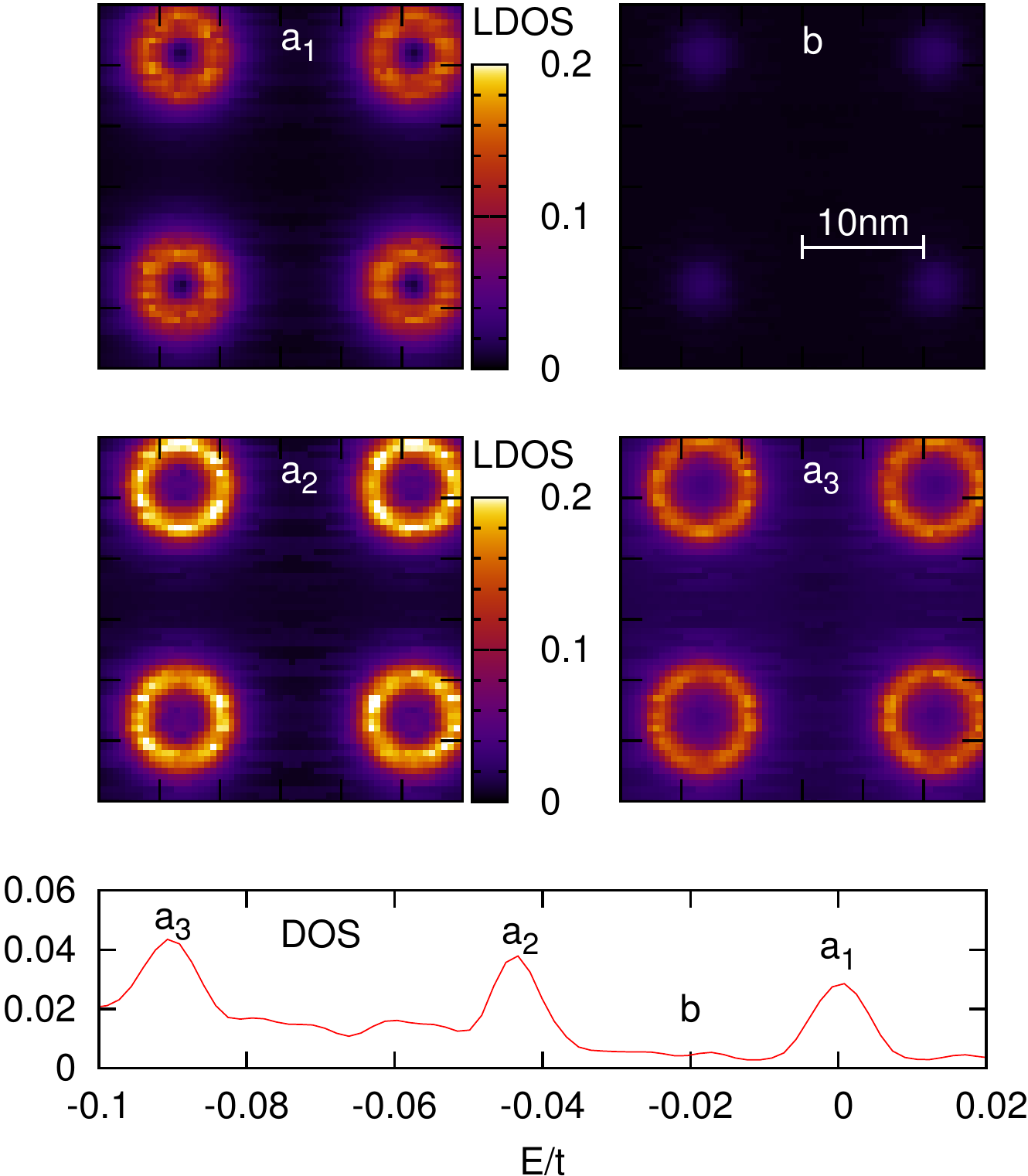}
\caption{(Color online) LDOS intensity plots for the central part of a (larger) square quantum-dot superlattice with 8$\times$8 dots (upper panels). 
The LDOS is given for the energies indicated in the lower panel, showing the mean DOS of the whole panel. 
To diminish finite-size effects the LDOS was calculated using 4096 Chebyshev moments only; 
therefore the splitting of the $a_m$ bands is not resolved here (recall that the resolution of the 
KPM scales with the inverse square root of the number of Chebyshev moments\cite{WWAF06}). 
System parameters are $R=4.775$nm, $D=19.1$nm, and $V/t=0.17092$ [as in Fig. \ref{fig2}~(d)].}
\label{fig3}
\end{figure}

To confirm the spatial localization of the states associated with the dot normal modes, we depict in Fig.~\ref{fig3} the LDOS for the four representative  energies indicated 
in the lower DOS panel. Here the first three energies fit to the corresponding dot modes, the fourth energy b gives an account of the 
situation in bulk graphene away from the resonances.  For the energies close to the dot-bound modes the LDOS profile is reflective of the quantum dot superlattice structure. Within the dot 
regions the intensity of the LDOS is enhanced in a ring-shape. Since the KPM has a finite energy resolution the LDOS  assembles contributions
from several eigenstates in the energetic vicinity of the target energy $E$.\cite{SSBFV10} 
We like to emphasize, however,  that these dot states are not strictly localized in real space. They are in 
superposition with each other (and also with bulk graphene states), leading to coherent transport but on a strongly reduced energy scale.   
For energies far off the resonances the LDOS is almost uniformly distributed, see panel energy b. Here the quantum dot superlattice behaves like a pristine graphene sample.

\begin{figure}[t]
\centering
\includegraphics[width=0.95\linewidth]{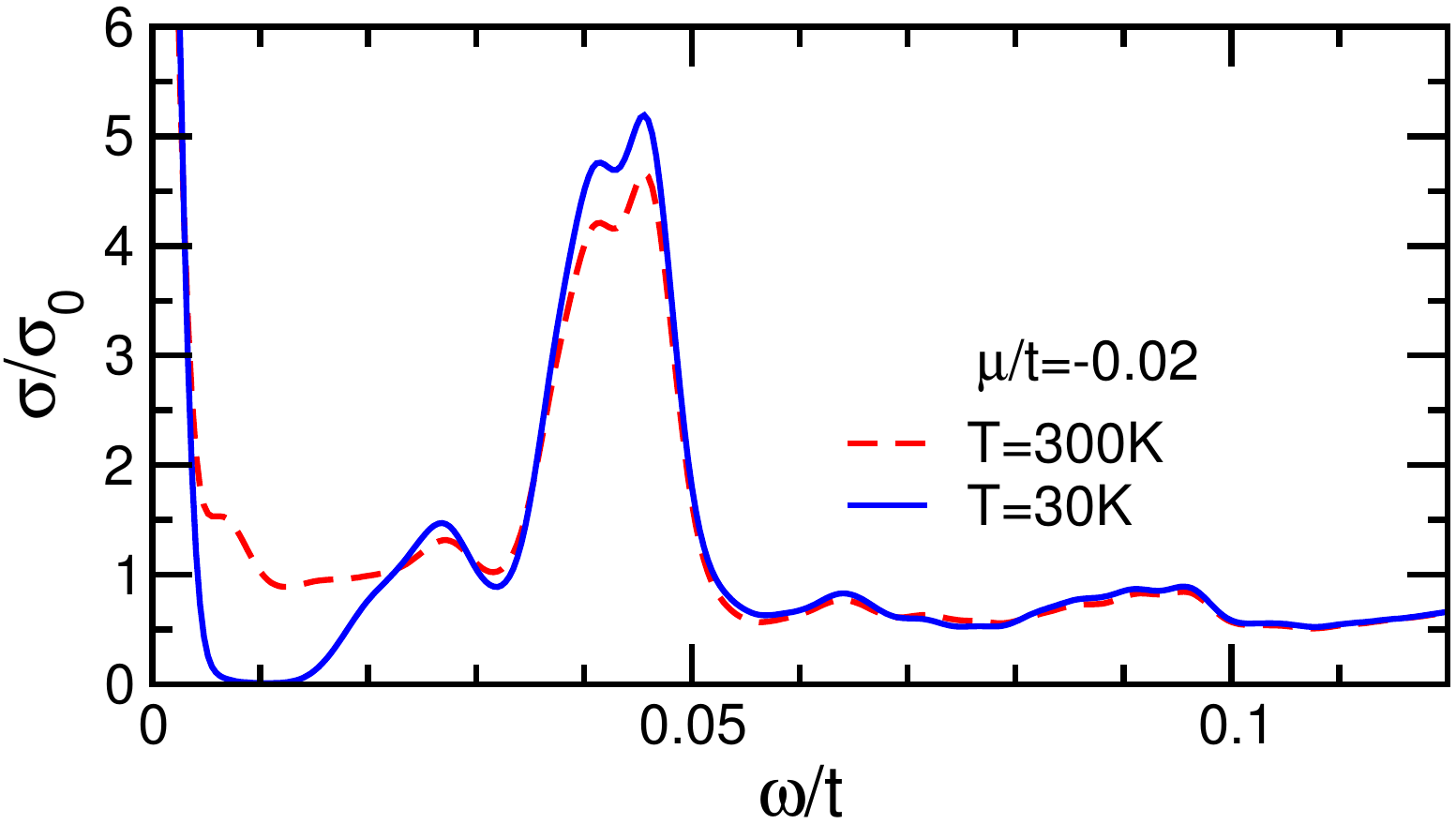}
\caption{(Color online) Optical conductivity (in units of $\sigma_0={\rm e}^2/8\hbar$)  for a 20$\times$20 graphene quantum dot superlattice with $V/t=0.17092$ and $R=4.775$nm, $D=19.1$nm. We chose $\mu/t=-0.02<0$ so that the Drude peak appears at $\omega=0$ and the transition between the occupied $a_2$ and the unoccupied $a_1$ mode appears at $\omega/t \simeq 0.045$ (note that the transition $a_3$ to $a_2$ does not appear as both modes are occupied). }
\label{fig4}
\end{figure}
We next demonstrate that transitions between the dot-bound states could be induced optically. To this end we have calculated the optical response 
of the graphene quantum dot array. Figure~\ref{fig4} gives the optical conductivity for the system studied in Figs.~\ref{fig2}~(d) and ~\ref{fig3}.   
Besides the Drude peak at $\omega=0$, noticeable absorption is only found for the transition from the $a_1$- to the $a_2$-band, which corroborates the optical selection rule
$a_m\to a_{m\pm1}$ obtained within the Dirac approximation.\cite{PHF13} Tuning $\mu$, different optical transitions might be singled out. 
As a matter of course the gap in the optical absorption spectrum fills with spectral weight at higher temperatures.  

\begin{figure}[t]
\centering
\includegraphics[width=\linewidth]{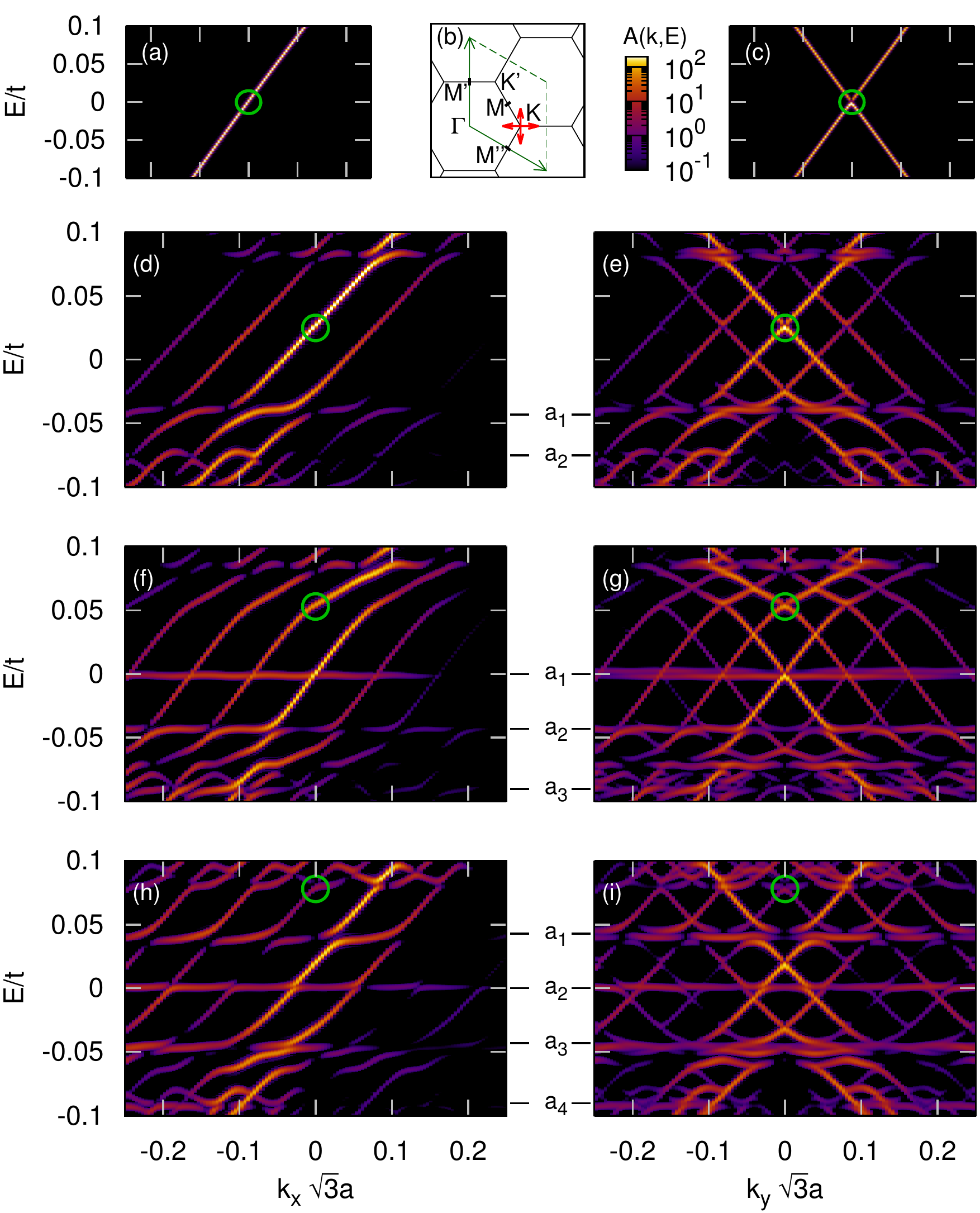}
\caption{(Color online)  Single-particle spectral function along the $\overline{\Gamma K}$ direction (horizontal; left-hand panels) and parallel to the $\overline{\Gamma M^\prime}$ direction (vertical, right-hand panels) through the Dirac  $K$ point, as 
indicated in the upper central figure (b), showing graphene's Brillouin zone. Panels (a)  and (c) give $A(\vec k,E)$ for a finite sample of pristine graphene ($V/t=0$) with PBC. Below, results for a 20$\times$20 dot-superlattice with $R=4.775$nm and  $D=19.1$nm are shown. In (d) and (e) $V/t=0.10727$ (mode $a_0$ falls on $E=0$), in (f) and (g) $V/t=0.17092$ (mode $a_1$ on $E=0$) and in (h) and (i) $V/t=0.22911$ (mode $a_2$ on $E=0$). The green marker (circle) traces the energy shift of the nodal point for pristine graphene when $V$ is increased. In view of the transfer of spectral weight to other nodal points it should, however, no longer be identified as the 'genuine' Dirac point. Note also that for mode $a_1$ at $E=0$ the Dirac point of pristine graphene evolves into the nodal point at $E=0$ when the dot spacing $D$ is reduced [compare panel (e) and Fig. \ref{fig6}~(b)]. Within the KPM 8192 moments were used.}
\label{fig5}
\end{figure}

\begin{figure}[t]
\centering
\includegraphics[width=\linewidth]{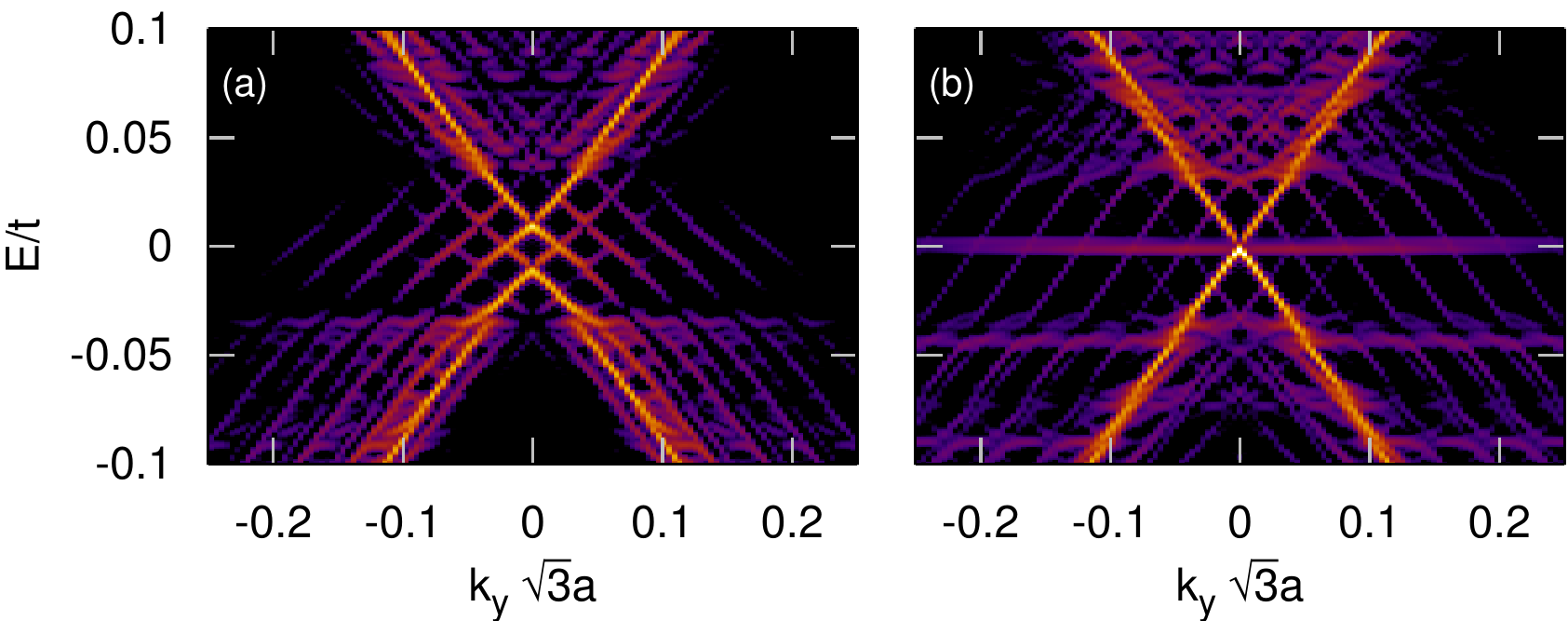}
\caption{(Color online)  Single-particle spectral function parallel to the $\overline{\Gamma M^\prime}$  direction through the Dirac $K$ point for a superlattice of 10$\times$10 dots with $R=4.775$nm and  $D=38.2$nm. In (a), $V/t=0.10727$ so that the mode $a_0$ falls on $E=0$ [as in Fig. \ref{fig5} (d) and (e)]. In (b), $V/t=0.17092$ so that mode $a_1$ falls on $E=0$ [as in Fig. \ref{fig5} (f) and (g)]. }
\label{fig6}
\end{figure}

To investigate the interplay of the dot-bound modes with superlattice effects we calculate the energy-momentum dependence of the single-particle spectral function $A(\vec k,E)$. Reflected in angle-resolved photoemission spectroscopy, this quantity gives insight into the  electronic band structure. 
Figure~\ref{fig5} displays the results in the vicinity of the Dirac point. In comparison to the perfectly linear energy bands 
of pure graphene [where $V_i\equiv 0$; see panels (a) and (c)], we observe for the graphene quantum dot array, on the one hand, a `ladder' of nearly dispersionsless bands formed by the quasi-bound dot states, and on the other hand, a sequence of  dispersive bands displaced against each other by reciprocal superlattice vectors. The latter bands will collapse if $D\to\infty$ (compare with Fig. \ref{fig6} showing results for a larger $D$). Note also the emergence of secondary nodal points due to intersection of energy bands at the edges of the superlattice Brillouin zone.

To assess how the dot-bound modes affect the dispersive bands of the superlattice we show $A(\vec k,E)$ in Fig. \ref{fig5} for different applied voltages at the dots. In panels (d) and (e) $V/t=0.10727$ so that the first $a_0$ mode is at $E=0$. This mode is not localized at the dots and no dispersionless band is formed at $E=0$. Instead, the mode $a_0$ hybridizes with the extended states outside the dot. This leads to a shift of the original Dirac cone to higher energy as propagating states also reside in the dots where the potential is higher. Furthermore, the group velocity at this nodal point is reduced by about 26\% which is even larger than  the  reduction by 19 \% obtained in second order perturbation theory.\cite{PYS08} In panels (f) and (g) $V/t=0.17092$ so that the mode $a_1$ falls on $E=0$. This mode is very sharp and shows only negligible hybridization with the propagating states outside the dot. Hence, the dispersion-less band originating from the $a_1$ mode is superimposed at $E=0$ on a Dirac cone  which is only marginally affected by the dots. The renormalization of the group velocity at the higher Dirac point amounts to 51\% in agreement with 48\% in second order perturbation theory. The different behavior between the modes $a_0$ and $a_1$ at $E=0$ is also reflected by Fig. \ref{fig6} where results for a larger $D$ are shown. For $a_0$ hybridization between electronic states inside and outside the dot transfers spectral weight to the newly emerging nodal points and the original Dirac point at $E=0$ vanishes. For the dot-localized $a_1$ mode, which couples negligibly to the extended states outside the dots, the original Dirac cone is preserved. Panels (h) and (i) of Fig. \ref{fig5} show the case of mode $a_2$ at $E=0$. This mode overlaps with the broad second $a_0$ mode so that the nodal point is shifted to higher $E$. We conclude that if the dots support only one very sharp mode $a_{m>0}$ at $E=0$ the dot superlattice leaves the conical energy dispersion of pristine graphene close to $E=0$ intact and the dispersionless dot band is merely superposed. This situation is best realized for the mode $a_1$.

\begin{figure}[t]
\centering
\includegraphics[width=0.9\linewidth]{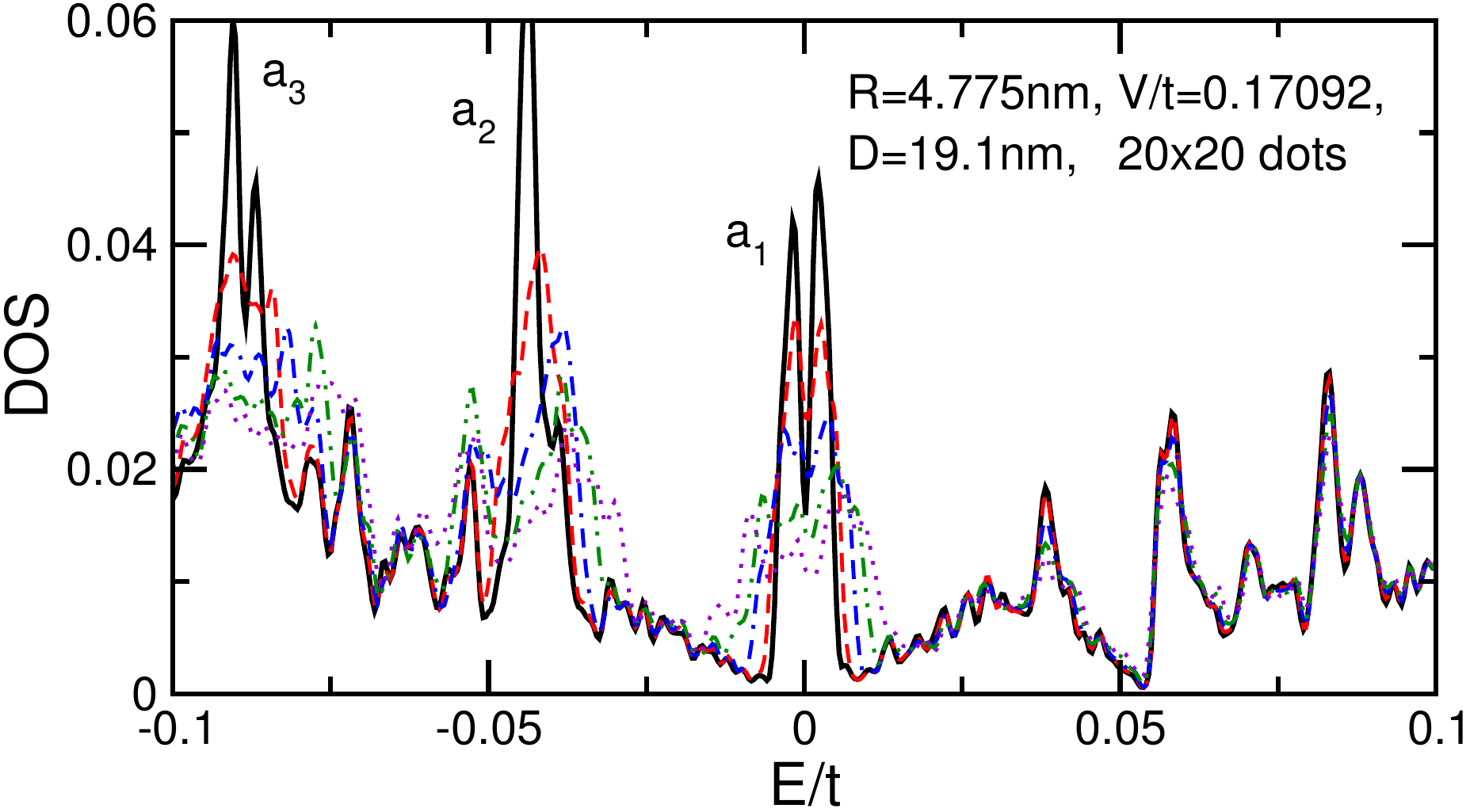}
\caption{(Color online) DOS for a 20$\times$20 superlattice sample of graphene quantum dots with random radii. 
Shown are the results for a single, typical realization, where all the dot radii were drawn from a uniform distribution of radii
with mean value $R=4.775$nm and $R_{(n,m)}/R\in$ [0.975, 1.025] (red dashed line), [0.95, 1.05] (blue dashed-dotted line),
[0.925, 1.075] (green double-dot-dashed line), and [0.9, 1.1] (violet dotted line). The black line gives the DOS without disorder. Again, we have $D=19.1$nm and $V/t=0.17092$.}
\label{fig7}
\end{figure}

\begin{figure}[t]
\centering
\includegraphics[width=0.98\linewidth]{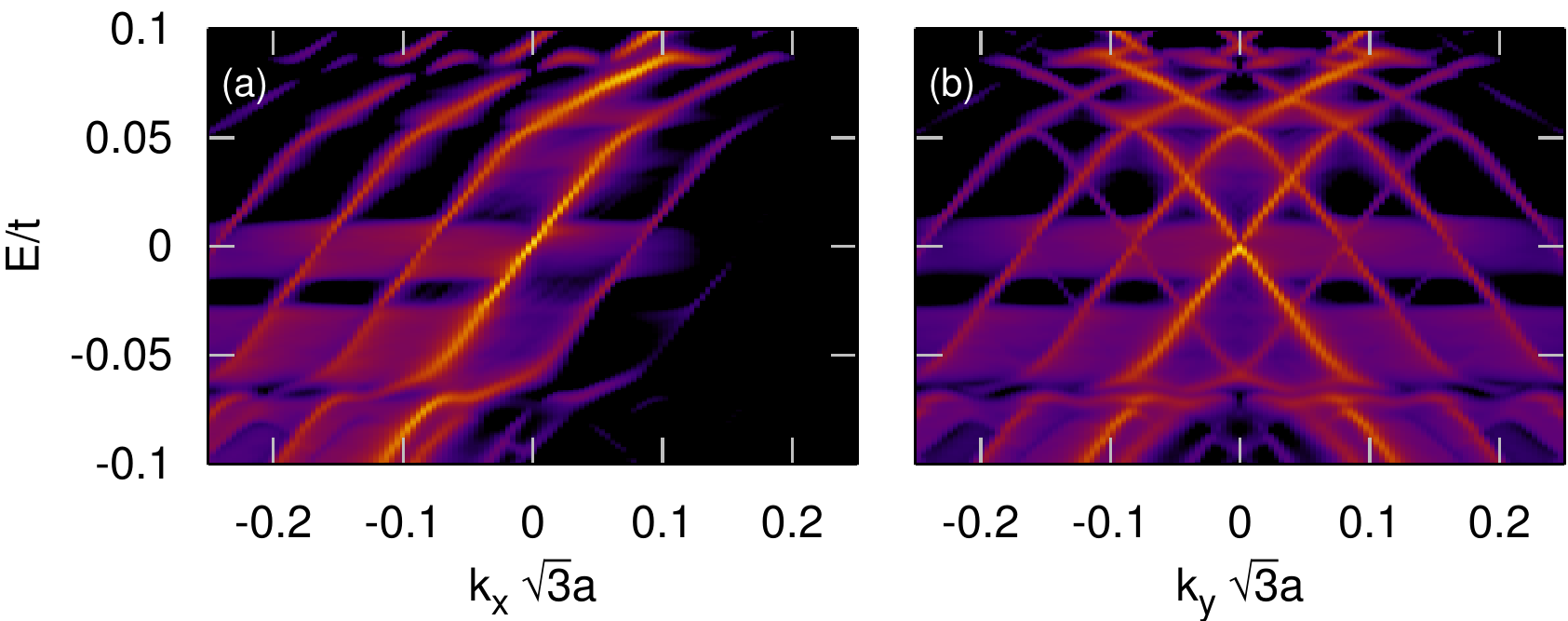}
\caption{(Color online) Single-particle spectral function through the Dirac $K$ point for a 20$\times$20 graphene quantum dot superlattice with random $R_{(n,m)}/R\in$ [0.9, 1.1]. The parameters $R$, $D$, and $V$ are chosen as in Figs.~\ref{fig5} (f) and 5~(g) so that the mode $a_0$ falls on $E=0$.}
\label{fig8}
\end{figure}

\begin{figure}[t]
\centering
\includegraphics[width=0.98\linewidth]{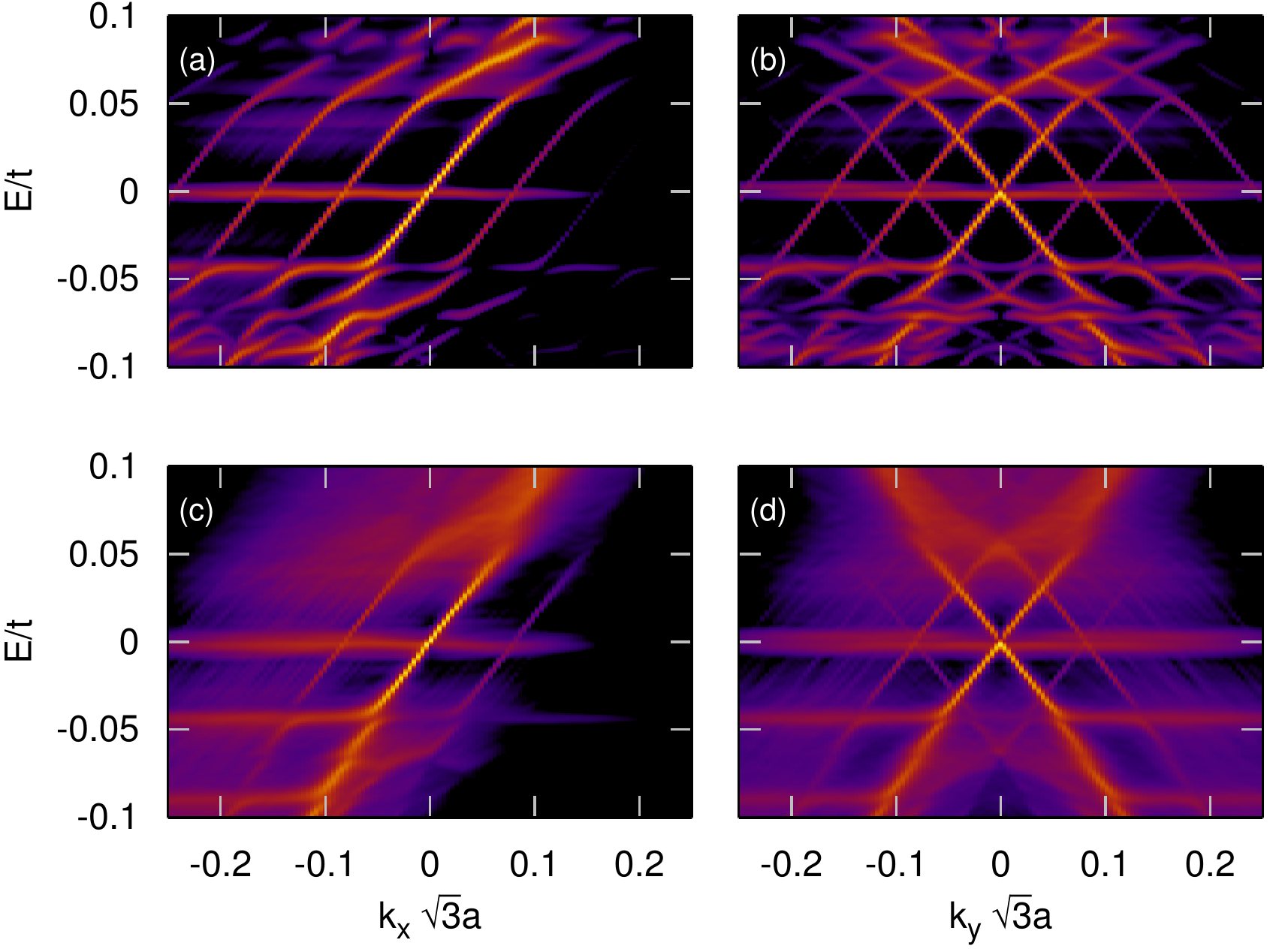}
\caption{(Color online) Single-particle spectral function through the Dirac $K$ point for a 20$\times$20 graphene quantum dot superlattice where the dots are displaced from their superlattice sites by $r_{(n,m)} \in [0,\Delta r]$  in random direction. In panels (a) and (b) $\Delta r=0.1$ in panels (c) and (d) $\Delta r =0.4$. The parameters $R$, $D$ and $V$ are chosen as in Figs.~\ref{fig5} (f) and 5~(g) so that the mode $a_0$ falls on $E=0$.}
\label{fig9}
\end{figure} 
 
Finally, we address the question of how disorder of a certain kind will affect the results shown so far. 
Intrinsic disorder, for instance, leads to the formation of electron-hole puddles \cite{MAU08} characterized by potential variations of typically less than 50 meV ($\approx$ 0.017$t$) which is small compared to the dot potential in our study. Hence, we expect our results to be relatively robust against intrinsic disorder and focus in the following on variations of the radii and the spacing of the gate-defined quantum dots as these should be uncertain to some extent in experiments.
Therefore we study, on the one hand,  a square superlattice of quantum dots with radii uniformly distributed around a mean value. Figure~\ref{fig7} presents the DOS of typical samples of such a random system. Obviously the peaks stemming from the quasi-bound states still exist but are considerably washed out if the disorder increases.  The same happens to the optical absorption (not shown). Looking at the single-particle excitation spectrum, we realize  that the almost dispersionsless bands, originating from the small overlap of the dot quasi-bound states, were  destroyed, see Fig.~\ref{fig8}. That means, quasi-bound dot states are still there but their coherence is lost.  The dispersive graphene states, on the contrary, are rather insusceptible  against the randomness induced by the different size of the quantum dots, at least close to the Dirac points $K$ and $K'$. If on the other hand, the dots are randomly shifted away from their superlattice sites, see Fig.~\ref{fig9}, the displaced dispersive bands become much weaker while the central cone as well as the dot-induced dispersion-less bands persist. We have also considered superlattices with elliptic dots (not shown). As the electron confinement is optimal only for circular dots non-circular dots lead to a significant broadening of the flat bands.

To conclude, superlattices of gate-defined quantum dots in graphene show clear indications of dot-bound modes in the (local) density of states, the optical conductivity, and the single particle spectral function. For superlattices with only one sharp localized mode at the charge neutrality point a dispersionless dot band emerges while the conical energy dispersion is preserved and pinned to $E=0$. For other choices of the dot potential the group velocity at the Dirac cone is significantly renormalized. Our results could be probed by angle-resolved photo-emission spectroscopy and scanning tunneling microscopy experiments and might guide the design of quantum dot superlattices in graphene.

\acknowledgements
This work was supported by the Deutsche Forschungsgemeinschaft through the priority programmes 1459 `Graphene' and 1648 `Software for Exascale Computing', and by the Center for Integrated Nanotechnologies at the Los Alamos National Laboratory  via DOE Contract No. DE-AC52-06NA25396.

\end{document}